\documentclass[12pt]{article} \textwidth=6in \oddsidemargin=0in
\usepackage{amssymb} 
\begin{document}
\def\ov{\over} \def\be{\begin{equation}} \def\cd{\cdots}
\def\ee{\end{equation}} \def\l{\ell} \def\iy{\infty}
\def\D{\Delta} \def\L{\Lambda} \def\ep{\epsilon}
\def\({\left(} \def\){\right)} \def\k{\kappa} \def\d{\delta} 
\def\e{\eta} \def\x{\xi} \def\o{\omega} \def\inv{^{-1}}
\def\s{\sigma} \def\ph{\varphi} \def\ps{\psi} \def\Z{\mathbb Z}
\def\ld{\ldots} \def\C{\mathcal C} \def\noi{\noindent}
\def\bc{\begin{center}} \def\ec{\end{center}} \def\qed{\hfill$\Box$}
\def\fr{\frac} \def\La{\Lambda} \def\({\left(} \def\){\right)}
\def\S{\mathcal S} \def\M{\mathcal M}

\hfill  October 31, 2013

\bc{\large\bf On the Diagonal Susceptibility
of the 2D Ising Model}\ec

\bc{\large\bf Craig A.~Tracy}\\
{\it Department of Mathematics \\
University of California\\
Davis, CA 95616, USA}\ec

\bc{\large \bf Harold Widom}\\
{\it Department of Mathematics\\
University of California\\
Santa Cruz, CA 95064, USA}\ec
\begin{quote}
\bc \textbf{Abstract} \ec
We consider the diagonal susceptibility of the  isotropic 2D Ising model for temperatures below
the critical temperature.  For a parameter $k$ related to temperature and the interaction constant, we extend
the diagonal susceptibility to complex $k$ inside the unit disc, and prove the conjecture that the unit circle is a
natural boundary. 
\end{quote}

\bc{\bf I. Introduction}\ec

For the 2D Ising model \cite{mccoy1, mccoy2, palmer},  after  the zero-field free energy \cite{onsager1} and the spontaneous magnetization \cite{onsager2, yang},  the most important zero-field thermodynamic quantity is the magnetic susceptibility $\chi$. 
Since the free energy is known only in zero magnetic field, the susceptibility is usually studied through its relation with the zero-field spin-spin correlation function:
\be \beta^{-1} \chi = \sum_{M,N\in\Z}\left\{ \langle \s_{0,0}\s_{M,N} \rangle - \M^2 \right\} \label{chi}\ee
where $\beta=(k_B T)^{-1}$, $T$ is temperature, $k_B$ is Boltzmann's constant and  $\M$ is the spontaneous magnetization.
If $T_c$ denotes the critical temperature, we recall that for the isotropic   2D Ising model, i.e.\ horizontal and vertical interaction constants have the same value $J$,
the spontaneous magnetization is given for $T<T_c$ by
\be \M=(1-k^2)^{1/8}\label{M}\ee
where $k:=\left(\sinh 2\beta J\right)^{-2}$ and $\M$ is zero for $T>T_c$. (For $0<T<T_c$ we have $0<k<1$.)

The analysis of $\chi=\chi(T)$ in the neighborhood of the critical temperature $T_c$ has a long history.   We refer the
reader to McCoy \textit{et al.} \cite{mccoy3} for a review of these developments.  The analysis of $\chi$ for
\textit{complex temperatures} was initiated by Guttmann and Enting \cite{guttmann} and by Nickel \cite{nickel1, nickel2}.   (For further
developments see \cite{chan, orrick}.) Nickel's analysis takes as its beginning
the (commonly called) form-factor or particle expansion of the spin-spin correlation function \cite{wmtb}.  For $T<T_c$ this expansion is an infinite sum whose $n$th summand is a $2n$-dimensional integral.  From an asymptotic analysis of these integrals, Nickel was led to conjecture that $\vert k \vert =1$ is a \textit{natural boundary} for $\chi$.  As Nickel himself noted, the analysis is nonrigorous  since one must show that there are no cancellations of singularities in the sum.  This has turned out to be a difficult problem to resolve rigorously.

In Boukraa \textit{et al.} \cite{boukraa1},  these authors,  building on results of \cite{lyberg},  introduce a simplified model for $\chi$, called the \textit{diagonal susceptibility} $\chi_d$, which is defined by having ``a magnetic field which acts only on one diagonal of the lattice.''  
(See \cite{assis} for further developments.) Thus the
analogue of (\ref{chi}) is
\be \beta^{-1}\chi_d=\sum_{N\in\Z}\left\{\langle \s_{0,0}\,\s_{N,N}\rangle-\M^2\right\} .\label{chid}\ee

In this paper  we consider $\chi_d$ only for  $T<T_c$, in which case $k<1$. Then we extend $\chi_d$  to $k$ complex 
with $|k|<1$.
Using the Toeplitz determinant representation of the diagonal correlations \cite{mccoy1, stephenson}, we first derive the known representation of  $\chi_d$ in terms of a sum of multiple integrals $\S_n$. The derivation is different from those in \cite{bugrii1, bugrii2}, \cite{lyberg} and \cite{witte}. As in \cite{witte} we use the identity of Geronimo-Case \cite{geronimo} and Borodin-Okounkov
\cite{borodin} relating a Toeplitz determinant to the Fredholm determinant of a product of Hankel operators. 
(For simplified proofs of  the GCBO formula, see \cite{basor, bottcher}.)
But here we go from there to the multiple integral representation directly using a general identity for the integral of a product of determinants \cite{andreief} (see eqn.~(1.3) in \cite{tw}).  For further background on the relationship between Toeplitz determinants and Ising correlations, see
 \cite{baxter, DIK}. 
 
In Section 4 we show that for each root of unity $\ep\ne\pm1$ a certain derivative of a certain $\S_n$ is unbounded as $k\to\ep$ radially, while the same  derivative of the sum of the other $\S_n$ remains bounded (Lemma 4).  Thus, the unit circle $\vert k \vert=1$ is a natural boundary for  
$\chi_d$.  This proves the conjecture by Boukraa \textit{et al.} \cite{boukraa1}.  We note that 
in \cite{boukraa1} the authors present an argument that the singularity of $\S_n$ at an $n$th root of unity $\ep$ is of the form
$(k-\ep)^{2n^2-1} \log (k-\ep)$. Lemma 2 in Section IV formalizes this statement and fills in details of the proof. 

\newpage
\bc{\bf II. Toeplitz determinant representation}\ec

It was shown in \cite{mccoy1, stephenson} that for $N>1$ the diagonal correlation has a representation as an $N\times N$ Toeplitz determinant:
\[\langle\s_{0,0}\,\s_{N,N}\rangle=\det\left(\ph_{m-n}\right)_{1\le m,n\le N}.\]
Here 
\[ \ph(\x)= \left[{1-k \x^{-1}\ov1-k \x}\right]^{1/2},\] 
and
\be\ph_m={1\ov2\pi i}\int\ph(\x)\,\x^{-m-1}\, d\x,\label{phm}\ee
with integration over the unit circle.
(We have $\langle\s_{0,0}^2\rangle=1$.)

As in \cite{witte} we invoke the formula of Geronimo-Case \cite{geronimo}  and Borodin-Okounkov \cite{borodin} to write the Toeplitz determinant in terms of the Fredholm determinant of a product of Hankel operators. We have $\ph(\x)=\ph_{+}(\x)\,\ph_{-}(\x)$, where
\[\ph_{+}(\x) = (1-k \x)^{-1/2}\>\>\>\textrm{and}\>\>\> \ph_{-}(\x) = (1 - k \x^{-1})^{1/2}.\]
Since $|k|<1$ these extend analytically inside and outside the unit circle, respectively. The square roots are determined by $\ph_+(0)=\ph_-(\iy)=1$. 

The Hankel operator $H(\ps)$ is the operator on $\l^2(\Z^+)$ with kernel $(\ps_{i+j+1})_{i,j\ge0}$,
where $\ps_m$ given in analogy with (\ref{phm}). The operator $H_N(\ps)$ has kernel $(\ps_{N+i+j+1})$.

Using $\ph_\pm(\x)=1/\ph_\mp(\x\inv)$, we find that the formula of GCBO gives
\[\det\left(\ph_{m-n}\right)_{1\le m,n\le N}=\M^2\,\det\Big(I-H_N\Big({\ph_-\ov\ph_+}\Big)
\,H_N\Big({\ph_+\ov\ph_-}\Big)\Big).\]
Thus, if we define
\be\La(\x)={\ph_-(\x)\ov\ph_+(\x)}=\sqrt{(1-k \x) (1-k/\x)}\,,\ \ \ \ K_N=H_N(\La)\,H_N(\La\inv),\label{LaK}\ee
then 
\[ \beta^{-1}\chi_d = 1-\M^2+2\M^2\sum_{N=1}^\iy\left[ \det(I-K_N)-1\right] = 1+\M^2(2\S-1), \]
where
\be \S=\sum_{N=1}^\iy \left[ \det(I-K_N)-1\right].\label{S}\ee

In what follows we extend $\La$ to be holomorphic in the complex plane cut along $[0,k]\cup[k\inv,\iy]$.
\pagebreak

\bc{\bf III. Formula for \boldmath$\S$}\ec

We use a slightly different notation for Hankel operators here.

\noi{\bf Proposition}. Let $H_N(du)$ and $H_N(dv)$ be two Hankel matrices acting on $\l^2(\Z^+)$ with $i,j$ entries
\be\int x^{N+i+j}\,d u(x),\ \ \ \int y^{N+i+j}\,dv(y),\label{uv}\ee
respectively, where $u$ and $v$ are measures supported inside the unit circle. Set $K_N=H_N(du)\,H_N(dv)$. Then
\[\sum_{N=1}^\iy [\det(I-K_N)-1]\]
\[=\sum_{n=1}^\iy{(-1)^n\ov (n!)^2}
\int\cd\int{\prod_i x_iy_i\ov1-\prod_i x_iy_i}\(\det\({1\ov 1-x_iy_j}\)\)^2\,\prod_i du(x_i)\,dv(y_i),\]
where indices in the integrand run from 1 to $n$.

\noi{\bf Proof}. The Fredholm expansion is
\[\det(I-K_N)=1+\sum_{n=1}^\iy{(-1)^n\ov n!}\sum_{p_1,\ld,p_n\ge0}\det(K_N(p_i,p_j)).\]
Therefore its suffices to show that
\[\sum_{N=1}^\iy\,\sum_{p_1,\ld,p_n\ge0}\det(K_N(p_i,p_j))\]
\[={1\ov n!}\int\cd\int{\prod_i x_iy_i\ov1-\prod_i x_iy_i}\(\det\({1\ov 1-x_iy_j}\)\)^2\,du(x_1)\cd du(x_n)\,dv(y_1)\cd dv(y_n).\]
We have
\[K_N(p_i,p_j)=\int\int{x^{N+p_i}\,y^{N+p_j}\ov1-xy}\,du(x)\,dv(y).\]
It follows by a general identity \cite{andreief} (eqn.~(1.3) in \cite{tw}) that
\[\det(K_N(p_i,p_j))={1\ov n!}\int\cd\int \det(x_i^{N+p_j})\,
\det(y_i^{N+p_j})\,\prod_i{1\ov 1-x_iy_i}\,\prod_i du(x_i)\,dv(y_i)\]
\[={1\ov n!}\int\cd\int \Big(\prod_i x_iy_i\Big)^N\,\det(x_i^{p_j})\,\det(y_i^{p_j})\,\prod_i{1\ov 1-x_iy_i}\,
\prod_i du(x_i)\,dv(y_i).\]
Summing over $N$ gives
\pagebreak

\[\sum_{N=1}^\iy\,\det(K_N(p_i,p_j))=\]
\[{1\ov n!}\int\cd\int {\prod_i x_iy_i\ov 1-\prod_i x_iy_i}\,\det(x_i^{p_j})\,\det(y_i^{p_j})\,\prod_i{1\ov 1-x_iy_i}\,\prod_i du(x_i)\,dv(y_i).\]
(Interchanging the sum with the integral is justified since the supports of $u$ and $v$ are inside the unit circle.)

Now we sum over $p_1,\ld,p_n\ge0$. Using the general identity again (but in the other direction) gives
\[\sum_{p_1,\ld,p_n\ge0}\det(x_i^{p_j})\,\det(y_i^{p_j})=
n!\,\det\(\sum_{p\ge0}x_i^p\,y_j^p\)=n!\,\det\({1\ov 1-x_iy_j}\).\]

We almost obtained the desired result. It remain to show that
\be\det\({1\ov 1-x_iy_j}\)\,\prod_i{1\ov 1-x_iy_i},\label{notquite}\ee
which we obtain in the integrand, may be replaced by
\be{1\ov n!}\,\(\det\({1\ov 1-x_iy_j}\)\)^2.\label{replace}\ee
This follows by symmetrization over the $x_i$. (The rest of the integrand is symmetric.) For a permutation $\pi$, replacing the $x_i$ by $x_{\pi(i)}$ multiplies the determinant in (\ref{notquite}) by sgn\,$\pi$, so to symmetrize we replace the other factor by
\[{1\ov n!}\sum_\pi {\rm sgn}\,\pi\,{1\ov 1-x_{\pi(i)}y_i}={1\ov n!}\,\det\({1\ov 1-x_iy_j}\).\]
Thus, symmetrizing (\ref{notquite}) gives (\ref{replace}).\qed 

We apply this to the operator $K_N=H_N(\La)\,H_N(\La\inv)$ given by (\ref{LaK}). The matrix for $H_N(\La)$ has $i,j$ entry
\[{1\ov2\pi i}\int\L(\x)\,\x^{-N-i-j-2}\,d\x,\]
where the integration may be taken over  a circle with radius in $(1,|k|\inv)$. Setting $\x=1/x$ and using $\L(1/x)=\L(x)$ we see that the entries of $H_N(\La)$ are given as in (\ref{uv}) with
\[du(x)={1\ov2\pi i}\,\L(x)\,dx,\]
and integration is over a circle $\C$ with radius in $(|k|,1)$.
Similarly $H_N(\La\inv)=H_N(v)$ where in (\ref{uv})
\[d\ps(y)={1\ov2\pi i}\,\L(y)\inv\,dy,\]
with integration over the same circle $\C$.

Hence the Proposition gives
\be \S=\sum_{n=1}^\iy \S_n,\label{sumS}\ee
where
\[\S_n={(-1)^n\ov (n!)^2}{1\ov(2\pi i)^{2n}}
\int\cd\int{\prod_i x_iy_i\ov1-\prod_i x_iy_i}\(\det\({1\ov 1-x_iy_j}\)\)^2\,\prod_i {\L(x_i)\ov \L(y_i)}\, \prod_i dx_i\,dy_i,\]
with all integrations over $\C$.

We deform $\C$ to the contour back and forth along the interval $[0,k]$, and then make the substitutions $x_i\to k x_i,\ y_i\to k y_i$. We obtain
\be \S_n={1\ov (n!)^2}\,{\k^{2n}\ov\pi^{2n}}\,\int_0^1\cd\int_0^1{\prod_i x_iy_i\ov1-\k^n\prod_i  x_iy_i}\;\(\det\({1\ov 1-\k x_iy_j}\)\)^2\,
\prod_i {\L_1(x_i)\ov \L_1(y_i)}\,\prod_i dx_i\,dy_i,\label{Sn1}\ee
where we have set
\[\k=k^2,\ \ \ \L_1(x)=\sqrt{{(1-x)(1-\k x)\ov x}}.\] 

Using the fact that the determinant in the integrand is a Cauchy determinant we obtain the alternative expression
\be \S_n={1\ov (n!)^2}\,{\k^{n(n+1)}\ov\pi^{2n}}\,\int_0^1\cd\int_0^1{\prod_i x_iy_i\ov1-\k^n \prod_i  x_iy_i}\;{\D(x)^2\,\D(y)^2\ov\prod_{i,j}(1-\k\,x_iy_j)^2}\,
\prod_i {\L_1(x_i)\ov \L_1(y_i)}\,\prod_i dx_i\,dy_i,\label{Sn2}\ee
where $\D(x)$ and $\D(y)$ are Vandermonde determinants.  Clearly, $\S_n$ is
holomorphic in $\k$ for $\vert\k\vert<1$.  It is straightforward to prove that the sum (\ref{sumS})
converges uniformly in $\k$ for $\vert \k \vert \le r$ for all $0<r<1$; and hence,  $\S$ is holomorphic in the
$\k$ unit disc.

\bc{\bf IV. Natural boundary}\ec

\noi{\bf Theorem}. The unit circle $|\k|=1$ is a natural boundary for $\S$.

There will four lemmas. In these, $\ep\ne1$ will be an $n$th root of unity and we consider the behavior of $\S$ as $\k\to\ep$ radially.

For $\l\ge0$ we use the representation (\ref{Sn2}) and look at
\be\int_0^1\cdots\int_0^1
{\prod_i x_iy_i\ov(1-\k^n \prod_i x_iy_i)^{\l+1}}\;{\D(x)^2\,\D(y)^2\ov\prod_{i,j}(1-\k\,x_iy_j)^2}\,\prod_i{\L_1(x_i)\ov\L_1(y_i)}\,\prod_i dx_i\,dy_i,\label{lint}\ee
where all indices run from 1 to $n$. This will be the main contribution to $d^\l \S_n/d\k^\l$.  

\noi{\bf Lemma 1}. The integral (\ref{lint}) is bounded when $\l<2n^2-1$ and it is of the order $\log(1-|\k|)\inv$ when $\l=2n^2-1$.

\noi{\bf Proof}. First we establish the first part of the statement. The numerator in the first factor is bounded and the denominator in the second factor is bounded away from zero as $\k\to\ep$ since $\ep\ne1$.

If $\prod_i x_iy_i<1-\d$ then the rest of the integrand is bounded except for the last quotient, and the integral of that is $O(1)$.

If $\prod_i x_iy_i>1-\d$ then each $x_i,\,y_i>1-\d$ and the integrand has absolute value at most a constant times
\[{\D(x)^2\,\D(y)^2\ov|\k^{-n}-\prod_i x_iy_i|^{\l+1}}\,
\prod_i\sqrt{{1-x_i\ov 1-y_i}}.\]

We assumed that $\k\to\ep$ along a radius, so $\k^{-n}>1$. Therefore we get an upper bound if we replace $\k^{-n}$ by 1. Then in the integral we make the substitutions $x_i=1-\x_i,\ y_i=1-\e_i$  (so $\x_i,\,\e_i<\d$) and we obtain 
\[{\D(\x)^2\,\D(\e)^2\ov(1-\prod_i (1-\x_i)(1-\e_i))^{\l+1}}
\,\prod_i \sqrt{\x_i\ov\e_i}.\]

Whenever $z_i\in[0,1]$ ($i=1,\ldots,m$) we have $z_1\cd z_{m}\le z_i$ for each $i$, and so averaging gives 
\[z_1\cd z_{m}\le\Big(\sum_j z_j\Big)/m,\]
and therefore
\[1-z_1\cd z_{m}\ge \sum_j (1-z_j)/m.\]
It follows that
\be 1-\prod_i (1-\x_i )(1-\e_i )\ge \sum_i(\x_i +\e_i)/2n.\label{lowerbound}\ee

Therefore the integrand above is at most $(2n)^{\l+1}$ times
\[{\D(\x )^2\,\D(\e )^2\ov(\sum_i(\x_i +\e_i ))^{\l+1}}
\,\prod_i \sqrt{\x_i\ov\e_i}.\]
This is homogeneous of degree $2n(n-1)-\l-1$. We first integrate over the region $\sum_i(\x_i+\e_i)=r$ and then over $r$. The resulting integral is at most a constant times
\[\int_0^{2n\d} r^{2n^2-\l-2}\,dr.\]
This is finite when $\l<2n^2-1$, and so the first statement of the lemma is established.
We note that the $(2n-1)$-dimensional volume of the region $\sum_i(\x_i+\e_i)=1$ is $1/\Gamma(2n)$, another nice factor which we can use if needed. But it won't be.

We now consider the integral when $\l=2n^2-1$. As before, the integral over the region $\prod_ix_iy_i<1-\d$ is $O(1)$, so we assume $\prod_ix_iy_i>1-\d$. In particular each $x_i,\,y_i>1-\d$. The factors $1-\k\,x_iy_j$ in the second denominator equal $1-\k(1+O(\d))=(1-\k)\,(1+O(\d))$ since $\k$ is bounded away from 1. From this we see that if we factor out $\k^{2n^3}$ from the first denominator and $(1-\k)^{n^2}$ from the second, the integrand becomes \[{\D(x )^2\,\D(y)^2\ov(\k^{-n}-\prod_ix_i y_i)^{2n^2}}\,
\prod_i\sqrt{{1-x_i\ov1-y_i}}\,(1+O(\d)).\]

We again make the substitutions $x_i=1-\x_i,\,y_i=1-\e_i$ and set $r=\sum_i(\x_i+y_i)$. Then since $\prod_i(1-\x_i)(1-\e_i)=1-r+O(r^2)$ the integrand becomes
\[{\D(\x )^2\,\D(\e)^2\ov(\k^{-n}-1+r+O(r^2))^{2n^2}}\,
\prod_i\sqrt{{\x_i\ov\e_i}}\,(1+O(\d)).\]   
The integration domain $\prod_ix_iy_i>1-\d$ becomes $r+O(r^2)<\d$, which is contained in $r<2\d$ and contains $r<\d/2$. The integral without the $O(\d)$ term is at least a constant times
\[\int_0^{\d/2}{r^{2n^2-1}\ov(\k^{-n}-1+2r)^{2n^2}}\,dr,\]
which is asymptotically a constant independent of $\d$ times $\log(\k^{-n}-1)\inv$ as $\k\to\ep$. Similarly the integral of the $O(\d)$ term is at most a constant independent of $\d$ times $\d\log(\k^{-n}-1)\inv$. Since $\d$ is arbitrarily small, this proves the lemma.\qed

\noi{\bf Lemma 2}. We have
\[\({d\ov d\k}\)^{2n^2-1}\S_n\approx \log(1-|\k|)\inv.\]

\noi{\bf Proof.} To compute the derivative of the integral in (\ref{Sn2}) one integral we get is a constant depending on $n$ times (\ref{lint}) with $\l=2n^2-1$. The other integrals are similar but in each the power in the denominator is less than $2n^2-1$ while we get extra factors obtained by differentiating the rest of the integrand for $\S_n$. These factors are of the form $(1-\k x_iy_i)\inv,\ (1-\k x_i)\inv$, or $(1-\k y_i)\inv$. By an obvious modification of the first statement of Lemma 1 we see that these other integrals are all bounded. The lemma follows.\qed

\noi{\bf Lemma 3}. If $\ep^m\ne1$ then 
\[\({d\ov d\k}\)^{2n^2-1}\S_m=O(1).\]

\noi{\bf Proof}. If $\ep^m\ne1$ all integrands obtained by differentiating the integral in (\ref{Sn2}) are bounded as $\k\to\ep$.\qed

\noi{\bf Lemma 4}. We have 
\[\sum_{m>n}\({d\ov d\k}\)^{2n^2-1}S_m=O(1).\]

\noi{\bf Proof}. We shall show that for $\k$ sufficiently close to $\ep$ all integrals we get by differentiating the integral for $S_m$ are at most $A^m\,m^m$, where $A$ is some constant. Note that the value of $A$ will change with each of its appearances. In may depend on $n$, but not on $m$.  Because of the $1/(m!)^2$ appearing in front of the integrals this will show that the sum is bounded.  

As before, we first use (\ref{Sn2}) (with $n$ replaced by $m$) and consider the integral we get when the first factor in the integrand is differentiated $2n^2-1$ times. All indices in the integrands now run from $1$ to $m$. 

First, 
\[|1-\k^m\prod_i x_iy_i|=|\k^m|\,|\k^{-m}-\prod_i x_iy_i|\ge
|\k|^m\,(1-\prod_i x_iy_i).\]
Next, $|1-\k\,x_i|\le 2$. Since $y_i\in[0,1]$ and $\k\in [0,\ep]$ we also have $\k y_i \in [0,\ep]$. Therefore $|1-\k y_i|\ge a$, where $a={\rm dist}(1,\,[0,\ep])$. Hence the integrand in (\ref{Sn2}) after differentiating the first factor has absolute value at most $A^m$ times
\be{1\ov(1-\prod x_iy_i)^{2n^2}}\;{\D(x)^2\,\D(y)^2\ov\prod_{i,j}|1-\k x_iy_j|^2}\prod_i\sqrt{{1-x_i\ov1-y_i}\;{y_i\ov x_i}}.\label{integrand}\ee
Since we also have  
\be|1-\k x_iy_j|\ge a,\label{a}\ee 
(\ref{integrand}) is at most
\[ a^{-m^2}\,{\D(x)^2\,\D(y)^2\ov(1-\prod x_iy_i)^{2n^2}}\,\prod_i\sqrt{{1-x_i\ov1-y_i}\;{y_i\ov x_i}}.\] 

With $x_i=1-\x_i,\,y_i=1-\e_i$ and $r=\sum_i(\x_i+\e_i)$ again, we first integrate over $r<\d$, where the small $\d$ will be chosen below. Using (\ref{lowerbound}) again, we see that the integrand is at most $A^m$ times
\[a^{-m^2}{\D(\x)^2\,\D(\e)^2\ov (\sum_i(\x_i+\e_i))^{2n^2}}\prod_i\sqrt{\x_i/\e_i}.\]
(The factor $m^{2n^2}$ coming from using (\ref{lowerbound}) and a bound for $\prod\sqrt{y_i/x_i}$ appearing in (\ref{integrand}) were absorbed into $A^m$.) When $\x_i,\,\e_i<1$ we have $\D(\x)^2,\,\D(\e)^2<1$, so integrating with respect to $r$ over $r<\d$, using homogeneity, gives at most a constant  times
\[a^{-m^2}\int_0^\d r^{2m(m-1)-2n^2+2m-1}\,dr=a^{-m^2}\int_0^\d r^{2m^2-2n^2-1}\,dr.\]
(The integral of the last factor over $r=1$ equals $(\pi/2)^m/\Gamma(2m)$.)
The integral is $O(\d^{2m^2})$ since $m>n$, and so the above is exponentially small in $m$ if we choose $\d^2<a$.

There remains the integral over the region $r>\d$, and for this we use the representation (\ref{Sn1}). We are led to (\ref{integrand}) with the second factor replaced by the absolute value of
\[\(\det\({1\ov 1-\k x_iy_j}\)\)^2.\]
From (\ref{lowerbound}) we see that in this region the first factor in (\ref{integrand}) is at most $(2m/\d)^{2n^2}$.
By (\ref{a}) and the Hadamard inequality the square of the determinant has absolute value at most $a^{-2m}\,m^m$. Therefore the integral over this region has absolute value at most
\[\({2m\ov\d}\)^{2n^2}\,a^{-2m}\,m^m\,\int_0^1\cd\int_0^1\prod_i\sqrt{{1-x_i\ov1-y_i}{y_i\ov x_i}}\;\prod_i dx_i\,dy_i.\]
The integral here is $A^m$, and so we have shown that the integral in the region $r>\d$ is at most $A^m\,m^m$.

This is a bound for only one term we get when we differentiate $2n^2-1$ times the integrand for $\S_m$. The number of factors in the integrand involving $\k$ is $O(m^2)$ so if we differentiate $2n^2-1$ times we get a sum of $O(m^{4n^2})$ terms. In each of the other terms the denominator in the first factor has a power no larger than $2n^2$ and at most $2n^2$ extra factors appear which are of the form $(1-\k x_iy_i)\inv,\,(1-\k x_i)\inv$, or $(1-\k y_i)\inv$. Each has absolute value at most $a\inv$, so their product is $O(1)$. It follows that we have the bound $A^m\,m^m$ for the sum of these integrals. Lemma~4 is established.\qed

\noi{\bf Proof of the Theorem}. Suppose $\k\to\ep$ radially, where $\ep\ne1$ is a root of unity. It is a primitive $n$th root of unity for some $n$. Then $\ep^m\ne1$ when $m<n$
so Lemma 3 applies for these $m$. Combining this with Lemma~4 and Lemma~2 gives
\[\({d\ov d\k}\)^{2n^2-1}\S\approx\log(1-|\k|)\inv.\]
Therefore $\S$ cannot be analytically continued beyond any such $\ep$, and these are dense in the unit circle.\qed

\noi{\bf Remark}. From the proofs of Lemma 3 and 4, with $2n^2-1$ replaced by $\l$ in both, one can see that $\S$ extends to a $C^\l$ function of $\k$ up to the boundary except for the $m$th roots of unity with $m\le\sqrt{(\l+1)/2}$\,. In particular, $\S$ extends to a function of class $C^6$ up to the boundary except for $\k=1$.

\begin{center}{\bf Acknowledgments}\end{center} 

The authors wish to thank Bernie Nickel and Barry McCoy for their helpful comments.
This work was supported by the National Science Foundation through grants DMS--1207995 (first author) and DMS--0854934 (second author).

\end{document}